\documentclass[draft]{agujournal2019}
\usepackage{url} 
\usepackage{lineno}
\usepackage{aas_macros}
\draftfalse

\journalname{Geophysical Research Letters}

\begin{document}

%
%


\title{Enhancement of impact heating in pressure-strengthened rocks in oblique impacts}

%
%




\authors{S. Wakita\affil{1,2}, H. Genda\affil{1}, K. Kurosawa\affil{3}, and T. M. Davison \affil{4}}


\affiliation{1}{Earth-Life Science Institute, Tokyo Institute of Technology, Meguro, Japan}
\affiliation{2}{Department of Earth, Environmental and Planetary Sciences, Brown University, Providence, RI, USA}
\affiliation{3}{Planetary Exploration Research Center, Chiba Institute of Technology, Narashino, Japan}
\affiliation{4}{Department of Earth Science and Engineering, Imperial College London, London, UK}




\correspondingauthor{Shigeru Wakita}{swakita@purdue.edu}




\begin{keypoints}
\item We studied heating in oblique impacts using a shock physics code
\item Oblique impacts can produce almost the same amount of heated mass as head-on impacts 
\item Various shock features could be produced by a single oblique impact event
\end{keypoints}

%
%


\begin{abstract}
Shock-induced metamorphism in meteorites informs us about the collisional environment and history of our solar system. 
Recently the importance of material strength in impact heating was reported from head-on impact simulations. 
Here, we perform three-dimensional oblique impact simulations, 
and confirm the additional heating due to material strength for oblique impacts. 
Despite a large difference in the peak pressure at the impact point at a given impact velocity, 
we find that the heated mass for an oblique impact is nearly the same as that for a head-on impact. 
Thus, our results differ from the previous finding
that the heated mass decreases as the impact becomes more oblique, and show that the additional shear heating is more effective for oblique impacts than for head-on impacts.
This also indicates that material ejected during oblique impact tends to experience lower shock pressures but higher temperatures.
\end{abstract}

%
%

%


%
%
%
%

\section{Introduction}
Since asteroids are thought to be surviving planetesimals or fragments of planetesimals \cite<e.g.,>{Morbidelli:2015aa}, 
they maintain primordial information about history of the solar system \cite<e.g.,>{DeMeo:2015aa}. 
Therefore, observation of current asteroids and analysis of meteorites originating from asteroids provide us with important keys to understanding the origin and evolution of the solar system.

Meteorites have been categorized depending on their degree of metamorphism, such as aqueous alteration and thermal metamorphism \cite<e.g.,>{Scott:2014aa}.
There are two major heat sources for their metamorphism.
One is an internal heating caused by the decay heat of short-lived radionuclides (e.g., $^{26}$Al), 
which follows thermal evolution of their parent bodies \cite<e.g.,>{Miyamoto:1982aa,Gail:2014aa,Wakita:2018aa}.
Another is an exogenous heating, such as impact heating.
Some meteorites contain unique textures which originate from impact events and are grouped by the degree of shock metamorphism  \cite{Stoffler:1991aa, Scott:1992aa, Rubin:1997aa, Stoffler:2018aa}.
Dehydrated minerals found in some chondrites are thought to be one line of evidence of impact heating \cite{Nakamura:2005aa,Nakato:2008aa,Abreu:2013aa},
because the temperature required for the dehydration is much higher than that for low degree of metamorphism in their host chondrites.
The short duration of the heating events for dehydrated minerals in parent bodies also supports impact heating as a reasonable heat source \cite{Nakato:2008aa}.
Given that we have an accurate understanding on the relationship between the impact conditions and the degree of heating, 
we can decode the impact events (such as the impact velocity and angle) from metamorphic features in meteorites.

Recently, \citeA{Kurosawa:2018aa} showed that the degree of impact heating had been underestimated.
\citeA{Melosh:2018aa} emphasized that the role of material strength in rocky materials have been overlooked for a long time. 
Plastic deformation of pressure-strengthened rocks due to shear strain against the material strength dissipates the kinetic energy of the materials in the shock-driven flow fields; this leads to a temperature rise during decompression.
The shock stages of meteorites typically grouped by their thermal properties \cite<e.g.,>{Stoffler:1991aa} may need to be reevaluated.
This additional heating could cause dehydration in chondrite parent bodies \cite{Wakita:2019aa}.
However, these pioneering studies only performed numerical calculations of head-on impacts with a two-dimensional shock physics code.
Oblique impacts occur more frequently than head-on collisions \cite{Shoemaker:1962aa,Kokubo:2010aa,Genda:2012aa}.
Therefore, it is necessary to explore whether material strength is still important in the case of oblique impacts.

Some numerical works have been done for oblique impacts. 
Methods for simulating oblique impacts include smoothed particle hydrodynamics \cite<SPH; e.g.,>{Monaghan:1992aa, Genda:2012aa} and grid-based codes such as CTH, SOVA and iSALE-3D \cite<e.g.,>{Pierazzo:2000aa, Pierazzo:2000ab, Elbeshausen:2009aa,Elbeshausen:2011aa,Elbeshausen:2013aa}. 
Previous work focused on the dependence of the crater volume and heated mass to the impact angle \cite{Pierazzo:2000aa, Pierazzo:2000ab, Elbeshausen:2009aa, Elbeshausen:2013aa, Davison:2011aa, Davison:2014aa}. 
The volume heated to any given temperature depends on the impactor diameter, mass, velocity and angle \cite{Pierazzo:2000aa} and, as a new finding in \citeA{Kurosawa:2018aa} and this work, the target strength.
\citeA{Davison:2014aa} showed that the heated mass decreases when the impact angle becomes shallower and highlighted the importance of target curvature when calculating the mass of heated material. 
\citeA{Davison:2014aa} estimated the post-shock temperatures using the peak shock pressures and assuming pure shock heating. 
 This does not account for the shear heating, which enhances the post-shock temperature as described in \citeA{Kurosawa:2018aa}.
Given this new result that the impact heating is enhanced when material strength is considered, 
here we perform a reevaluation of the findings of \citeA{Davison:2014aa}, 
where we simulate the impact for longer and can thus track the full temperature evolution of the material as in \citeA{Kurosawa:2018aa}. 
We extend the results of \citeA{Kurosawa:2018aa} by considering the effects of impact obliquity on impact heating.

Here, we report numerical simulations of planetesimal collisions using the iSALE-3D shock physics code. 
We compare the results between oblique and head-on collisions, 
and discuss the importance of material strength.

\section{Methods}
We perform impact simulations of a spherical projectile onto a flat-surface target  
by using the iSALE-3D shock physics code \cite{Elbeshausen:2009aa,Elbeshausen:2011aa,Collins:2016aa}. 
This code uses a solver as described in \citeA{Hirt:1974aa} and includes a strength model and a porous-compaction model \cite{Collins:2004aa,Melosh:1992aa,Ivanov:1997aa,Wunnemann:2006aa,Collins:2011aa}.
We consider the size of impactor ($R_{\rm imp}$) as  2 km, and the impact velocity ($v_{\rm imp}$) of 5 km/s,
the typical velocity in the current main asteroid belt \cite{Bottke:1994aa,Farinella:1992aa}.
We consider the impact angles ($\theta_{\rm imp}$) for a head-on collision (90$^\circ$) and an oblique collision (45$^\circ$).
The most probable $\theta_{\rm imp}$ of a randomly incident projectile is 45$^\circ$ \cite<e.g.,>{Shoemaker:1962aa}, which we  chose as a fiducial value.  

We use a strength model commonly used for geologic materials \cite{Collins:2004aa}, 
using the input parameters shown in \citeA<Table S1 of>{Kurosawa:2018aa},
and use the ANEOS equation of state for dunite \cite{Benz:1989aa} for both the projectile and target. 
Further details of the strength model are described in Supporting Information Text S1.
The damaged friction coefficient ($\mu$) in the model is one of the most important parameters for determining the heating of internal materials \cite{Kurosawa:2018aa,Wakita:2019aa}.
Therefore, we perform the models with and without material strength to clarify the role of the material strength on the degree of impact heating, using the fiducial value $\mu$ of 0.6 \cite<e.g.,>{Johnson:2015aa}.

We use Lagrangian tracer particles which are placed in each cell at the beginning of the calculation, and allowed to move around the Eulerian grid to track the history of the material. 
They can record both the instantaneous and peak values of pressure and temperature. 
Note that entropy is not currently calculated in iSALE-3D. The difference between the calculations in this and previous works is that we need to run our simulations for longer to capture both the initial shock heating and the subsequent shear heating; we need to use the temperature field rather than deriving temperatures from the peak pressures.
Validation of our numerical results, a resolution study, and a comparison of our methods with previous works are given in Supporting Information (Figures S1, S2, Texts S2, and S3).

\section{Results}

\begin{figure}
\noindent\includegraphics[width=\textwidth]{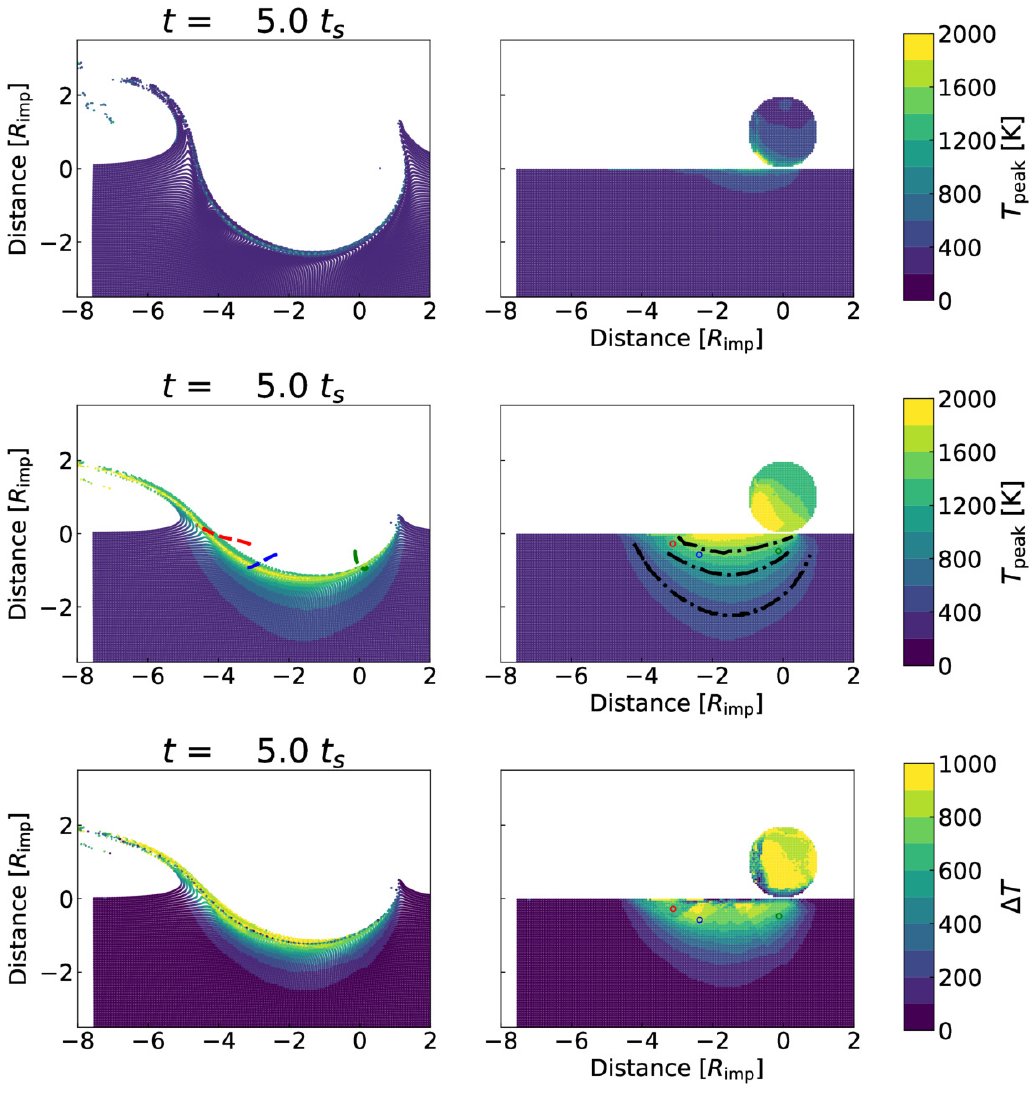}
\caption{Snapshots of a cross section at the impact point (0,0), in the plane that includes the impact trajectory, at 5 $t_s$ for the oblique impact case (45$^\circ$).
The color scale in the top and middle panels represents $T_{\rm peak}$, and in the bottom panels represents $\Delta T$ (see text).
The top panels depict the case without material strength and the middle and bottom panels with material strength. 
The left-hand panels show the material in its position at 5~$t_s$, while the right-hand panels are provenance plots that display the material mapped back to its original pre-impact position. 
The dash-dotted lines in the middle-right panel represent isothermal lines ($T_{\rm peak}$ = 500 K, 1000 K, and 1500 K).
The colored dashed lines in the middle-left panel depict the trajectories of selected tracer particles,
and their original positions are shown as open circles in the middle and bottom right panels.
}
\label{fig1}
\end{figure}

Figure \ref{fig1} shows distributions of peak temperature ($T_{\rm peak}$) at 5 $t_s$ ($t_s$ is a characteristic time for projectile penetration, $t_s$ = 2$R_{\rm imp}/v_{\rm imp}$) 
in the cases of a 45$^\circ$ impact with (middle and bottom rows) and without (top row) material strength. 
The case with material strength experiences much higher $T_{\rm peak}$ than the case without material strength (the pure hydrodynamic case).
According to the strength effects, $T_{\rm peak}$ for the case with the material strength increases during decompression as shown in \citeA{Kurosawa:2018aa}. 
In order to confirm this, we also show the temperature differences ($\Delta T$) between the peak temperatures and 
the temporal temperatures when it reaches its peak pressure in Figure \ref{fig1} (bottom panels).
We can see two highly heated ($\Delta T > \sim$ 1000 K) regions in the target (bottom right panel in Figure \ref{fig1}): 
one of them is near the impact point and the other is about $3R_{\rm imp}$ away from it and lies in the downrange direction of the incident trajectory of the projectile. 
The former heats up first (see Figure S3), then the heating of the latter area occurs. 
This wider heated region is different from head-on impacts, as shown below.

\begin{figure}
\noindent\includegraphics[width=\textwidth]{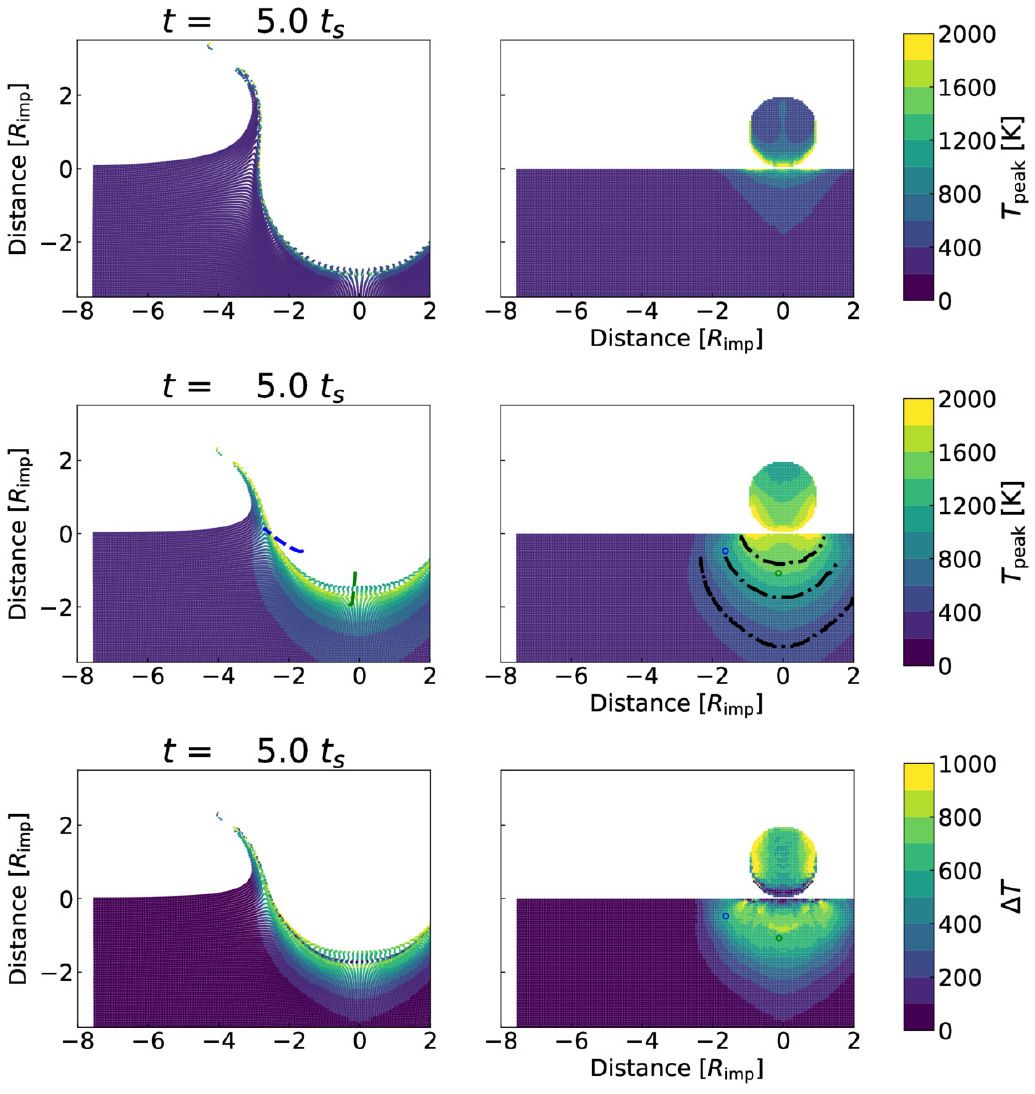}
\caption{Same as Figure \ref{fig1}, but for the case of head-on impact (90$^\circ$).}
\label{fig2}
\end{figure}

In order to compare the heated areas between head-on impacts (90$^\circ$) and oblique impacts (45$^\circ$),
we show the results of two head-on impacts in Figures \ref{fig2} and S4 (with and without material strength).
We confirmed that material strength enhances the heating for in head-on impacts in Cartesian coordinates (iSALE-3D), 
which is consistent with the results obtained in cylindrical coordinates (iSALE-2D) employed by \citeA{Kurosawa:2018aa} (see Text S4).
The highly heated region in the target is symmetrically distributed just below the impact point for the head-on impact, 
whereas it is shifted towards the downrange side for the oblique impact.
The heated region in the head-on impact is deeper than in the oblique impact. 

\begin{figure}
\noindent\includegraphics[width=\textwidth]{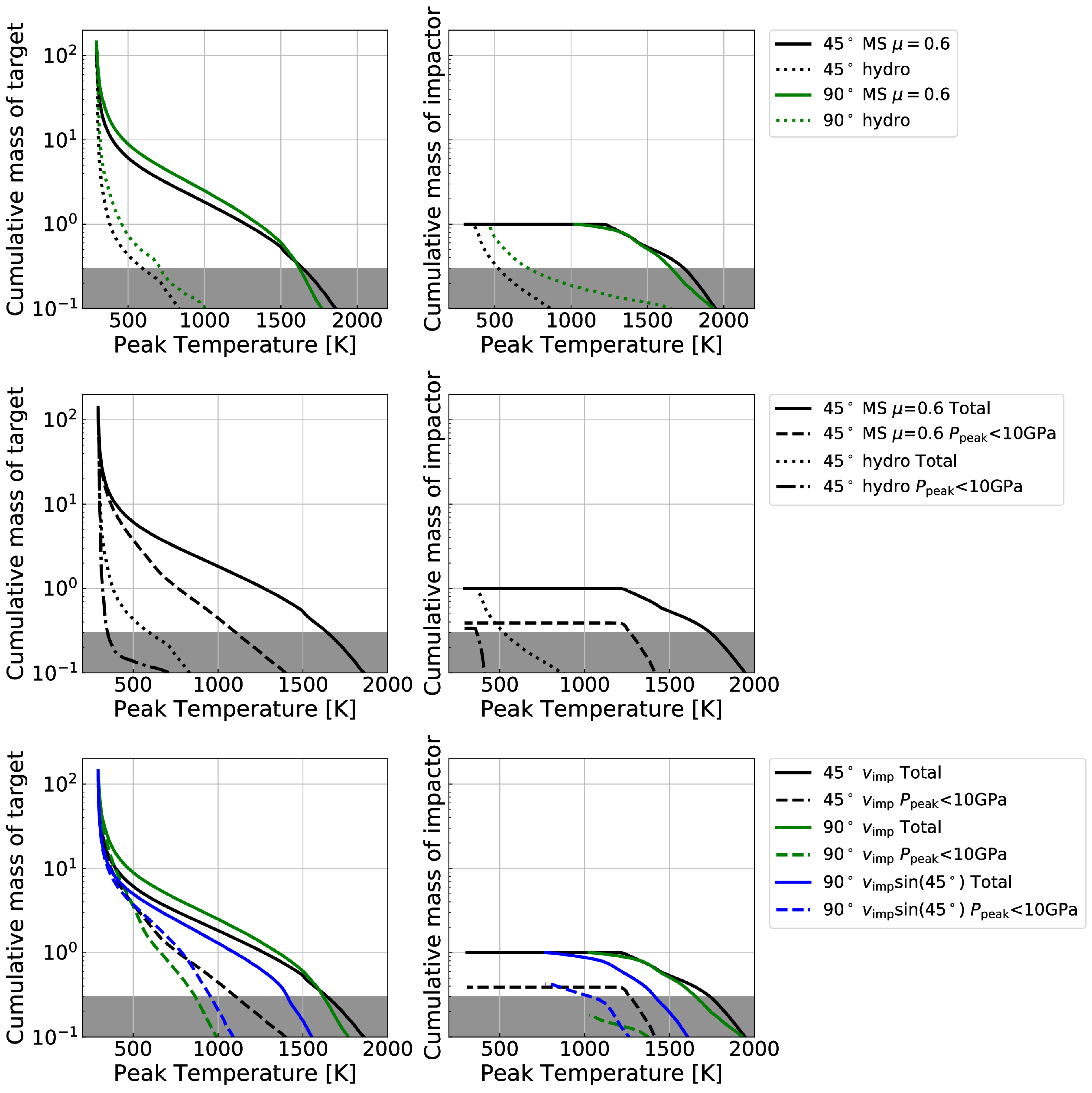}
\caption{Cumulative mass of $T_{\rm peak}$ normalized by $M_{\rm imp}$. 
Left panels depict results from the target, and right panels are for the impactor.
Top panels represent the case with and without material strength for oblique and head-on impacts (MS denotes the case with material strength and hydro does without that).
Middle and bottom panels represents the result of corresponding peak pressure conditions and $v_{\rm imp}$ (see legends).
The shaded region represents the artificial overhead region due to the overshooting of temperature (see Text S1).
}
\label{fig3}
\end{figure}

Figure \ref{fig3} shows the cumulative mass of target and impactor normalized by the impactor mass ($M_{\rm imp}$) according to $T_{\rm peak}$. 
Although these masses are calculated at the time of 5 $t_s$,
we have confirmed that these results do not significantly change after this time (see Figure S5). 
The difference in cumulative mass between the two cases with material strength and $\theta_{\rm imp}$ of 45$^\circ$ and 90$^\circ$ is less than a factor of two (top panels of Figure \ref{fig3}).  
This tendency is the same in both the target and impactor. 
Consequently, an oblique impact could produce a similar amount of heated material as a head-on impact with same $v_{\rm imp}$. 

The results of pure hydrodynamic cases are also shown as the dotted lines in the top panels of Figure \ref{fig3}. 
The amount of the heated material without material strength is much smaller than that in the cases with material strength. 
This clearly indicates the significance of material strength in impact heating
(see also Text S5 and Figure S6).
Therefore, material strength is an important factor which must be considered when calculating heating in oblique impacts.

\begin{figure}
\noindent\includegraphics[width=\textwidth]{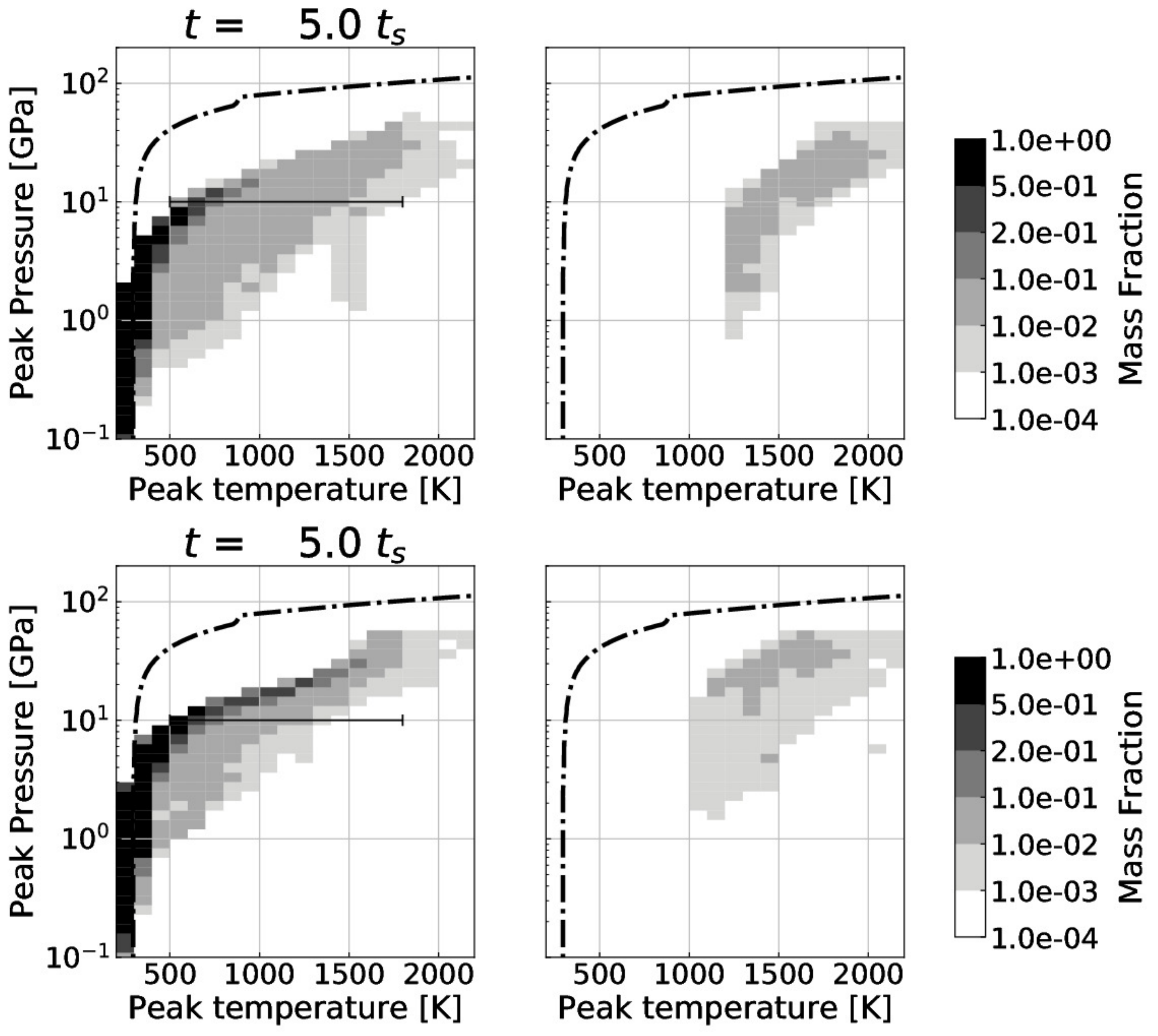}
\caption{Heatmaps of $T_{\rm peak}$ and $P_{\rm peak}$ at the time of 5 $t_s$. 
The gray contour represents their fraction which is normalized by $M_{\rm imp}$.
Top panels depict results of oblique impact (45$^\circ$), bottom ones are head-on impact (90$^\circ$) with material strength. 
Left panels represent outcomes from the target, and right panels are the impactor. 
Hugoniot curves for dunite are also shown as dash-dotted lines in each panel.
}
\label{fig4}
\end{figure}

Considering the cases with material strength we can examine how the relationship between peak temperature and peak pressure ($P_{\rm peak}$) changes between head-on and oblique impacts. 
Figure \ref{fig4} shows the distribution of $T_{\rm peak}$ and $P_{\rm peak}$. 
The peak temperature are systematically higher than the temperatures on the Hugoniot curve (dash-dotted lines) as discussed in previous works \cite{Kurosawa:2018aa, Wakita:2019aa}. 
The relationship between $T_{\rm peak}$ and $P_{\rm peak}$ differs in both impact cases. 
In the oblique impact (top left panel in Figure \ref{fig4}), the target can achieve a peak temperature in the range 1000 - 1500 K at a peak pressure of $\sim$ 10 GPa. 
One of the key differences between oblique and head-on collisions is that the temperature field appears to continue evolving for longer in the oblique case; for example, Figures~\ref{fig4} and S7 show that the mass fraction at peak pressure of 10~GPa in the peak temperature range of 1000 - 1500 K increases between 1.3~$t_s$ and 5~$t_s$,
while in head-on impact that mass fraction remains nearly the same between those two times.

\begin{figure}
\noindent\includegraphics[width=\textwidth]{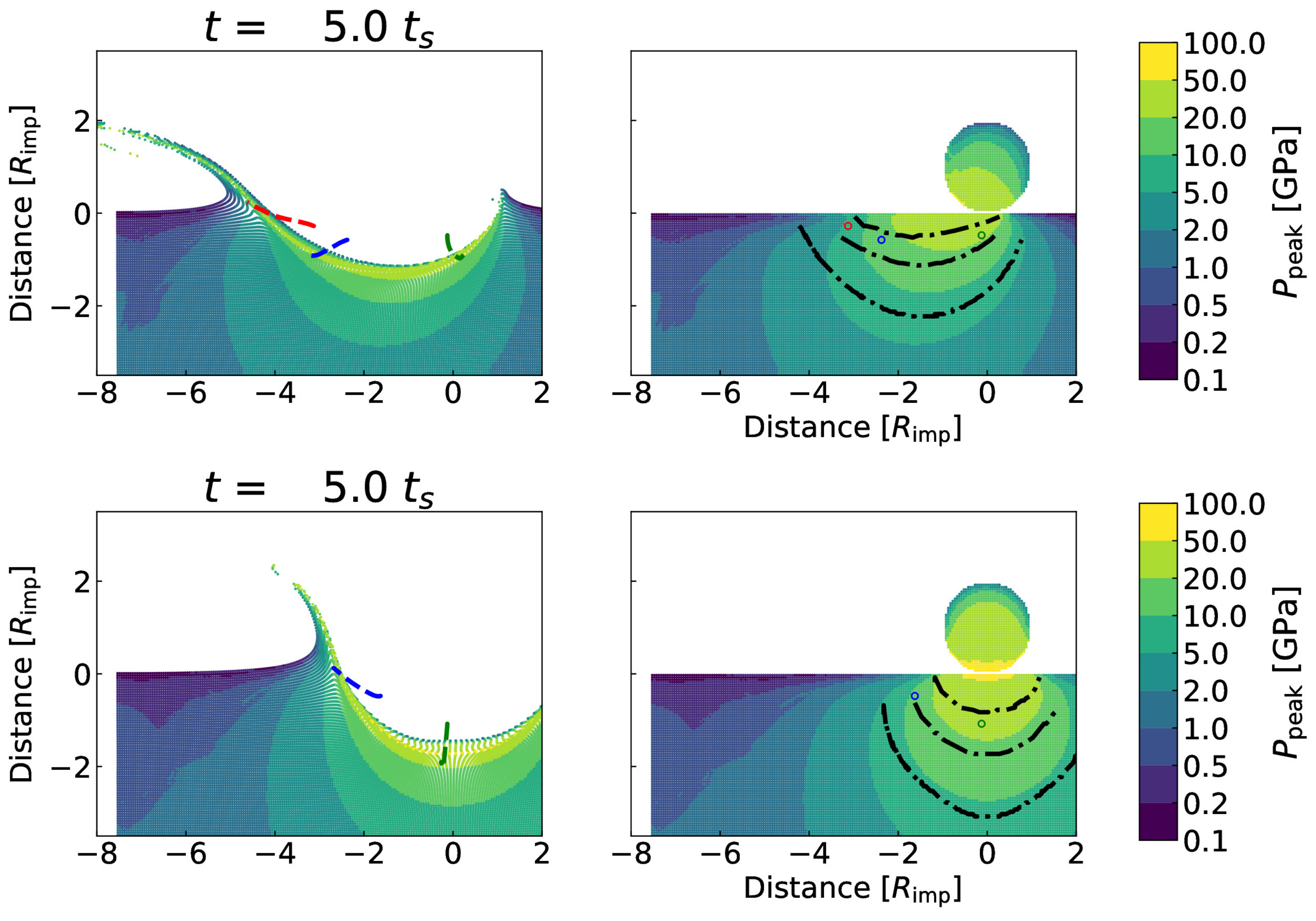}
\caption{Same as middle panels in Figures \ref{fig1} and \ref{fig2}, but color contours represent $P_{\rm peak}$. 
Top panels represent oblique impact (45$^\circ$) and bottom panels are head-on (90$^\circ$).
}
\label{fig5}
\end{figure}

We also checked the relationships of $P_{\rm peak}$ and $T_{\rm peak}$ from different points of view.
The cumulative mass relative to $T_{\rm peak}$ for $P_{\rm peak} <$ 10 GPa are plotted in middle and bottom panels of Figure \ref{fig3}.
Since $T_{\rm peak}$ in the case without material strength are reached by pure shock heating, for material with $P_{\rm peak} < $ 10~GPa, 
the temperature remains almost at the initial temperature.
In contrast, in the case with material strength, some material exceeds $T_{\rm peak} =$ 1000 K even at $P_{\rm peak}$ $<$ 10~GPa due to the contribution of shear heating (see the dashed lines in middle panels of Figure \ref{fig3}).
Thus, the enhanced heating at $P_{\rm peak} <$ 10 GPa is additional evidence of the importance of material strength in oblique impacts (see Text S5).

Figure \ref{fig5} shows the spacial distribution of $P_{\rm peak}$, overlain with isothermal lines.
This pressure distribution is similar to that found in previous work investigating impact heating \cite{Pierazzo:2000ab}.
As we can see in bottom panels of Figure \ref{fig3}, the heated mass under 10 GPa in the oblique impact is much larger than that in the head-on impact (compare the black and green dashed lines).
In the bottom-right panel of Figure \ref{fig5}, the 1000~K isotherm (the middle of the three dot-dashed isotherms) lies entirely within the 10--20~GPa contour on the head-on impact. However, in the oblique impact, the 1000~K isotherm crosses the 10~GPa contour in the downrange region of the target---i.e. some material which experienced $P_{\rm peak} <$ 10~GPa experienced temperatures above 1000~K. The red open circle on the top-right panel of Figure \ref{fig5} is an example of some of the material which experienced these P-T conditions.

When we focus on the highly heated region ($\Delta T > \sim$ 1000 K) in the bottom panels of Figures \ref{fig1} and \ref{fig2}, 
the area appears near the impact point in both $\theta_{\rm imp}$ cases. 
The other area, $3R_{\rm imp}$ away from the impact point, is observed only in the oblique case.
This area is produced by a further increase in volumetric strain due to the downrange movement of a distorted projectile, while this area is less shocked.
 An adjacent area ($\sim 4R_{\rm imp}$), which is only seen in oblique impacts, corresponds to the region which can reach $T_{\rm peak} > $ 1000 K at $P_{\rm peak} < $ 10 GPa (see Figures \ref{fig4} and \ref{fig5}).
This is because that shear deformation due to movement of the projectile produces additional heating for a wider area, such as the red colored tracer particle (see Figures \ref{fig1}, \ref{fig5}, and S4).
While the volume of heated material near the impact point is smaller in the oblique impact, 
this additional downrange heating means that the total mass of material heated is similar between the oblique and normal incidence angle impacts.

\section{Discussion}

We discuss the relationship between $\theta_{\rm imp}$ and $v_{\rm imp}$. 
It has been sometimes thought that the amount of heating in an oblique impact with $v_{\rm imp}$ could be approximated 
by a head-on impact with the vertical velocity component of $v_{\rm imp}$.
The vertical component of $v_{\rm imp}$ in oblique impacts with $\theta_{\rm imp}$ can be given as $v_{\rm imp}\sin(\theta_{\rm imp})$. 
\citeA{Pierazzo:2000aa} showed that the peak pressure in oblique impacts could be approximated by that produced by vertical impacts at $v_{\rm imp}\sin(\theta_{\rm imp})$ without material strength. 
Although a further correction is needed in the melt volume; 
the deviation in the melt volume produced by oblique impacts from the prediction by $v_{\rm imp}\sin(\theta_{\rm imp})$ model is within a factor of two.
Rarefaction waves travel faster in materials with strength than in hydrodynamic materials, meaning the shock wave will decay sooner.
In order to check this result when material strength is included, we also perform a head-on impact with $v_{\rm imp}\sin(45^\circ)$ (= 3.54 km/s; see blue lines in bottom panels of Figure \ref{fig3}). 
When we compare the results with oblique impact with $v_{\rm imp}$, the cumulative mass differs by a factor of three. 
Therefore, it is hard to reproduce the outcomes related to peak temperature from oblique impacts of $v_{\rm imp}$ by head-on impacts of $v_{\rm imp}\sin(\theta_{\rm imp})$,
when we include the effect of material strength.
This indicates that the peak temperature cannot be directly determined  from the peak pressure and their relationship is complicated.
This shows that the additional heating is effective in oblique impacts due to the material strength.
Previous studies on oblique impacts found that
the crater volume and heated mass could be correlated well with $\theta_{\rm imp}$ \cite<e.g.>{Elbeshausen:2009aa,Davison:2011aa,Davison:2014aa}.
Although the simulations in \citeA{Davison:2014aa} also included material strength with the parameters for weak rock, 
that work used a different method to obtain temperature after decompression which did not account for shear heating.
A more extensive suite of simulations over a wide range of impact angles is required to investigate the complex interaction of the shock and rarefaction waves in materials with strength, and derive a similar function which accounts for the shear heating described in this work.

Shear heating has implications for ejected material.
Our results indicate that oblique impacts produce nearly the same heated mass as head-on impacts and 
suggest that oblique impacts can generate a heated region with lower peak pressures than in a head-on impact. 
We infer that the ejecta from an oblique impact might also experience higher peak temperatures than a head-on impact at a given peak pressure,
although higher-resolution simulations are required to resolve the ejected materials in detail \cite<e.g.,>{Johnson:2014aa,Kurosawa:2018ab}.

All meteorites might experience one impact event, at least when they are excavated from their parent bodies. 
\citeA{Stoffler:1991aa} presented a useful tool to categorize meteorites into shock stages, depending on the degree of shock metamorphism. 
Our results show that oblique impacts could produce the same thermally-driven metamorphic properties for lower shock pressures than previously reported.
Thus, although a more thorough investigation of the effects of oblique impacts is necessary,
we propose that care is needed to decode shock temperatures experienced by meteorites, since there is not a direct link between peak shock pressure and temperature.

\section{Conclusion}
Oblique impacts are ubiquitous events on all solid bodies in the solar system. 
We numerically modelled oblique impacts of 45$^\circ$ at 5 km/s, which is a typical impact velocity in the main asteroid belt, with a strength model for rocky materials, and examined the peak temperature and peak pressure after the impacts. 
We confirmed the importance of material strength for additional heating in oblique impacts as well as head-on impacts that was shown in previous works.
We also found that the amount of heated mass is almost the same in head-on and oblique impacts.  
On the contrary, there is a difference in the experienced maximum temperature for peak pressures less than 10 GPa: 
a head-on impact does not reach 1000 K, but in an oblique impact it is possible to exceed 1000 K. 
The relationship between peak pressure and peak temperature reveals that oblique impact can generate a moderately heated region under lower peak pressure than head-on impacts.
The area heated by oblique impacts is located shallow and wide in the target, which is achieved in the downrange direction of the impactor's trajectory.
These enhanced heating processes in oblique impacts are due to a combination of the material strength and movement of the projectile.

\acknowledgments
We gratefully acknowledge the developers of iSALE-3D, including Dirk Elbeshausen, Kai W{\"u}nnemann, and Gareth Collins (http://www.isale-code.de). 
All our data are given by using iSALE-3D and our input files are available in Supporting Information Datasets S1--S3 and on website (https://doi.org/10.5281/zenodo.3515505). 
Please note that usage of the iSALE-3D code is restricted to those who have contributed to the development of iSALE-2D, 
and iSALE-2D is distributed on a case-by-case basis to academic users in the impact community.
It requires a registration from the iSALE webpage (http://www.isale-code.de) and usage of iSALE-2D and computational requirements are also shown in there.  
We directly plot figures from our binary data using pySALEPlot which is included in iSALE-3D and developed by TMD. 
A python script for middle panel of Figure 1 is available in Datasets S4.
Please also note that pySALEPlot in the current stable release of iSALE-2D (Dellen) would not work for the data from iSALE-3D.
We also thank Kai W{\"u}nnemann and an anonymous referee for their kind comments and suggestions.
Numerical computations were partially carried out on the analysis servers at the Center for Computational Astrophysics, National Astronomical Observatory of Japan. 
This work have been supported in part by JSPS, Japan KAKENHI Grant Number JP17H06457 and JP17H02990.
K.K. is supported by JSPS KAKENHI grant numbers JP17H01176, JP17H01175, JP17K18812, JP18HH04464, and JP19H00726. 
TMD is funded by STFC grant ST/S000615/1.


%
%


%
%
%
%
%

\end{document}